\documentclass[twocolumn,aps,prb,floatfix,superscriptaddress]{revtex4-2}
\usepackage{amsmath}
\usepackage{amsfonts}
\usepackage{amssymb}
\usepackage{graphicx}
\usepackage{soul}
\usepackage{mathtools}
\usepackage{color}
\usepackage{bm}
\usepackage{subfigure}
\usepackage{wasysym} 
\usepackage{graphicx}
\usepackage{dcolumn}
\usepackage[final]{hyperref} 
\hypersetup{
	colorlinks=true,       
	linkcolor=blue,        
	citecolor=blue,        
	filecolor=magenta,     
	urlcolor=blue         
}

\newcommand{\ii}{\text{i}}

\newcommand{\be}{\begin{equation}}
\newcommand{\ee}{\end{equation}}
\newcommand{\bea}{\begin{eqnarray}}
\newcommand{\eea}{\end{eqnarray}} 

\newcommand{\mb}{\mathbf}
 \newcommand{\new}[1]{\textcolor{black}{#1}}

 \makeatletter
\newsavebox{\@brx}
\newcommand{\llangle}[1][]{\savebox{\@brx}{\(\m@th{#1\langle}\)}%
  \mathopen{\copy\@brx\kern-0.5\wd\@brx\usebox{\@brx}}}
\newcommand{\rrangle}[1][]{\savebox{\@brx}{\(\m@th{#1\rangle}\)}%
  \mathclose{\copy\@brx\kern-0.5\wd\@brx\usebox{\@brx}}}
\makeatother
 
\begin{document}

\title{Visons in Kitaev Spin Liquids with Majorana Fermi Surfaces} 

\author{Caio V. S. Soares}
\affiliation{Departamento de F\'{i}sica Te\'{o}rica e Experimental,
Universidade Federal do Rio Grande do Norte, 59078-970 Natal, RN, Brazil}

\author{Rodrigo G. Pereira}%
\affiliation{Departamento de F\'{i}sica Te\'{o}rica e Experimental,
Universidade Federal do Rio Grande do Norte, 59078-970 Natal, RN, Brazil}
\affiliation{International Institute of Physics, Universidade\\
Federal do Rio Grande do Norte, 59078-970, Natal, RN, Brazil}%

\date{\today} 

\begin{abstract}
The excitation spectrum of  Kitaev quantum  spin liquids consists of itinerant  Majorana fermions, which can be gapless or gapped,  and vortices of a $\mathbb{Z}_2$ gauge field,  known as visons, which are \new{gapped within a stable $\mathbb Z_2$ spin liquid phase}. In this work, we  investigate visons in Kitaev-type models where the Majorana fermions form a Fermi surface. In this case, the   creation of a vison pair is analogous to introducing a local impurity potential in a metal. Since the gapless modes lead to strong finite-size effects, we compare the numerical calculation of the \new{two-vison gap} on finite lattices with the result from an analytical approach based on Green's function techniques. We find that the \new{two-vison gap} decreases as the size of the Fermi surface increases,  signalling  an instability of the quantum spin liquid ground state. We also  show that   larger Fermi surfaces tend to suppress the change in local spin correlations due to the Majorana-vison scattering potential.

\end{abstract}

\maketitle

\section{Introduction}

Visons are flux excitations of $\mathbb Z_2$ gauge fields that emerge in  fractionalized phases of matter \cite{Senthil2000,sachdev2023quantum}. The interplay between visons and other fractional excitations is crucial for the unique properties of $\mathbb Z_2$ quantum spin liquids (QSLs) \cite{lee2008end,savary2016quantum,zhou2017quantum}, with most studies  so far focusing  on gapped QSLs.  For example, in Anderson's resonating valence bond state \cite{anderson1973resonating,Kivelson1987}, the nontrivial mutual statistics between visons  and spinons  is responsible for the  topological ground-state degeneracy  \cite{sachdev2023quantum}. In the  non-Abelian phase of the   Kitaev honeycomb model (KHM)  in a uniform magnetic field \cite{KitaevAOP2006}, visons bind Majorana zero modes that behave as  Ising anyons.  The proposal that Kitaev interactions can be realized in  Mott insulators with strong spin-orbit coupling \cite{jackeli2009mott,chaloupka2010kitaev},  such  as $\alpha\text{-RuCl}_3$ \cite{plumb2014alpha,banerjee2016proximate,do2017majorana,takagi2019concept}, has motivated  the steady progress in the study of  Kitaev spin liquids in the past two decades \cite{winter2017models,Motome2020,matsuda2025kitaev}.

In stark contrast with gapped $\mathbb Z_2$ QSLs, Majorana fermions  in Kitaev spin liquids can also form gapless phases  known as Majorana metals. Despite being charge insulators, these systems  display power laws  in specific heat and thermal conductivity akin to  Fermi-liquid behavior \cite{barkeshli2013gapless,tikhonov2010quantum,o2016classification,pereira2018gapless,Zhang2019,bauer2019symmetry,Lee2023,Chari2021,Oliviero2024,Zhu2025,feng2026magneticfieldinducedphenomena}. Starting from the KHM at zero magnetic field, where the Majorana fermions have a Dirac spectrum, it is possible to obtain a Majorana Fermi surface by applying  a staggered magnetic field  induced by proximity with a N\'eel-ordered two-dimensional  magnet \cite{Takikawa2019,Nakazawa2022,yokoyama2025}. A concrete proposal to realize this effect is a heterobilayer of $\alpha$-RuCl$_3$ and the easy-axis antiferromagnet MnPS$_3$ \cite{consoli2026}.  A Majorana Fermi surface has also been conjectured to explain the observation of   incommensurate Friedel-like oscillations        in the local density of states near defects in $\alpha$-RuCl$_3$ \cite{Kohsaka2024}.

While  visons  have  been shown to play an important role in correlations in the  KHM \cite{baskaran2007exact,lahtinen2008spectrum,lahtinen2014perturbed,knolle2014dynamics,Joy2022},  little attention has been paid to  visons in  QSLs with Majorana Fermi surfaces. A basic property in this context  is the \new{two-vison gap}, which must remain finite within the $\mathbb Z_2$ QSL phase. Traditionally, the \new{two-vison gap} in the KHM has been evaluated numerically on finite-size lattices \cite{KitaevAOP2006,pedrocchi2011physical,zschocke2015physical,o2016classification}. When the Majorana fermions are gapped, such calculation converges exponentially fast with increasing  system size. However,   strong finite-size effects are expected  in the case of a Majorana Fermi surface.  It is important to determine  the magnitude of the \new{two-vison gap}   because a vanishing    gap    signals an instability of the $\mathbb Z_2$ QSL, either  toward magnetically ordered phases \cite{Knolle2018,Zhang2021,Sodemann2023,Chen2025} or other QSL phases \cite{Hickey2019}, thus constraining the parameter space where Majorana Fermi surface states can be observed.  Moreover, flux excitations with small gaps can affect thermodynamic and transport properties of QSLs at low but finite temperatures \cite{Nasu2015}.

In this work, we compute the energy gap for creating two adjacent visons in Kitaev-type models with Majorana Fermi surfaces \cite{yao2009algebraic,baskaran2009exact,chua2011exact,lai2011power}.  The main question is how the size of the Majorana Fermi surface affects vison properties. In addition to exact diagonalization on finite lattices, we employ  a recently developed analytical approach, valid in the thermodynamic limit, that relates the  \new{two-vison gap}   to a scattering   phase shift for Majorana fermions    \cite{panigrahi2023analytic}.  As our first example, we consider the KHM   perturbed by a three-spin interaction that  mimics  the effect of a staggered magnetization. Our second example is  the \new{Chua-Yao-Fiete} (CYF) model  for  spins $3/2$ on the kagome lattice \cite{chua2011exact,lai2011power}.  In both models, the size of the Fermi surface can be controlled by tuning a parameter that moves the Dirac cone in the Majorana fermion dispersion away from zero energy. 
 As the area enclosed by the Majorana Fermi surface increases, we find that the \new{two-vison gap} develops oscillations as a function of system size.  Moreover, the asymptotic value   decreases  and the gap eventually closes for a sufficiently  large Majorana Fermi surface. We connect this observation with  the effect of the Majorana Fermi surface on   local  spin correlations in the vicinity of the visons.

The  paper is organized as follows. In Sec. \ref{sec:model}, we present the models and their solution in terms of Majorana fermions coupled to $\mathbb Z_2$ gauge fields. We also discuss how the size of Majorana Fermi surface varies as we tune a control parameter. Section \ref{sec:visongap} contains our results for  the \new{two-vison gap}. We discuss the oscillations as a function of system size and, when possible, compare the numerical results for finite lattices with the analytical prediction in the thermodynamic limit.  We analyze  the effect of visons on   local spin correlations in Sec. \ref{sec:correl}.  Finally, we offer some concluding remarks in Sec. \ref{sec:concl}.

\section{Models \label{sec:model}}

In this section, we describe two    spin models that can be solved in terms of Majorana fermions coupled to  static $\mathbb Z_2$ gauge fields and yield  QSLs with Majorana Fermi surfaces. We start with the spin-1/2 KHM perturbed by three-spin interactions, and then turn  to the spin-$3/2$ CYF model on the kagome lattice.

\subsection{Perturbed Kitaev honeycomb model}

\begin{figure}
    \centering
    \subfigure[]{\includegraphics[width=0.26\textwidth]{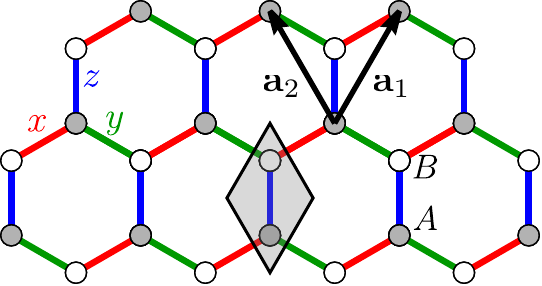}\label{lattice}} 
   \hspace{0.5cm}
    \subfigure[]{\includegraphics[width=0.13\textwidth]{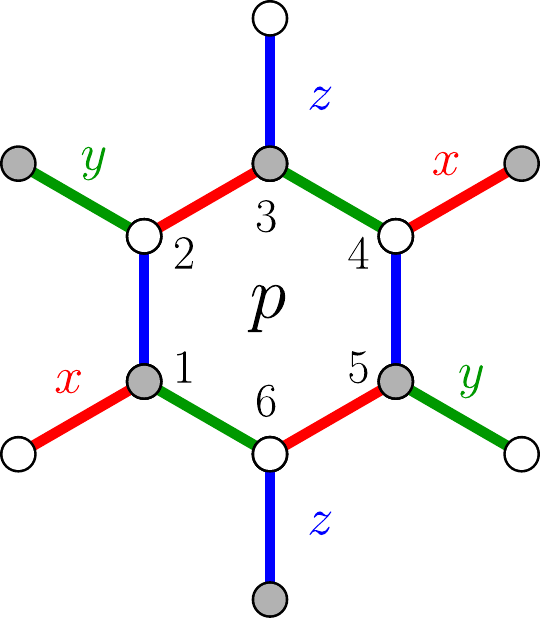}\label{plaquettekitaev}}
    \caption{Kitaev honeycomb model. (a) Sites in sublattices A and B  are represented by  gray and white dots, respectively. The three sets of nearest-neighbor bonds are labeled by $\alpha=x,y,z$, highlighted in red, green, and blue, respectively. The primitive lattice vectors are $\mb a_1=(\frac12,\frac{\sqrt3}2)$ and $\mb a_2=(-\frac12,\frac{\sqrt3}2)$.  (b) A plaquette $p$ containing six sites numbered from 1 to 6.  The conserved   quantity  associated with this plaquette is the operator  $W_p=\sigma_1^x\sigma_2^y\sigma_3^z\sigma_4^x\sigma_5^y\sigma_6^z$.}
    \label{f1}
\end{figure}

The KHM describes  spins $1/2$ on a honeycomb lattice that  interact via the Hamiltonian \cite{KitaevAOP2006}
\begin{equation}
H_K=-\sum_{\alpha=x,y,z}J_{\alpha}\sum_{\langle i,j\rangle_\alpha}\sigma_{i}^{\alpha}\sigma_{j}^{\alpha},
\end{equation}
where $J_\alpha$ are coupling constants and  $\sigma_i^\alpha$ are Pauli operators acting in the local Hilbert space at  site $i$. The nearest-neighbor bonds connecting sites in sublattices A and B are labeled by $\alpha=x,y,z$ as shown in Fig. \ref{lattice}.  
This model has an extensive set of mutually commuting  conserved  plaquette operators, 
$    W_{p}=\prod_{i\in p}\sigma_{i}^{\alpha_{i}}$, where the indices $\alpha_i$ for sites around the plaquette are chosen as shown  in Fig.  \ref{plaquettekitaev}.  As a consequence, the Hilbert space splits  into sectors labeled by the eigenvalues $w_p=\pm1$ of  all $W_p$. Writing $w_p=e^{i\varphi_p}$ with $\varphi_p\in\{0,\pi\}$, we interpret $\varphi_p$ as $\mathbb{Z}_2$ fluxes through the plaquettes. A theorem by Lieb \cite{lieb1994flux}   guarantees that the ground state of the pure KHM lies in the sector with $\varphi_p=0$ $\forall p$, which we call the flux-free sector.


We perturb the KHM by adding  the \new{three-spin interaction \cite{Takikawa2019,Nakazawa2022,yokoyama2025} }\be
H_{\rm pert}=-\sum_{\alpha}J'_\alpha\sum_{\langle i,j\rangle_\alpha\langle j,k\rangle_\beta } (-1)^{s_i} \sigma^\alpha_i\sigma^\gamma_j\sigma^\beta_k, \label{pertH}
\ee
where $s_i=0$ for $i\in \text{A}$ and $s_i=1$ for $i\in \text{B}$, and  $(\alpha,\beta,\gamma)$ is a cyclic permutation  of $(x,y,z)$. Note that the    sum runs over triangles with three adjacent sites $(i,j,k)$ forming two nearest-neighbor bonds labeled by $\alpha$ and $\beta$. The three-spin operator can be  regarded as a spin chirality operator. Importantly, the perturbation in Eq.  (\ref{pertH}) commutes with all plaquette operators. Moreover, it differs from the term that opens a gap in the non-Abelian  QSL phase \cite{KitaevAOP2006} because here the sign of the induced spin chirality  alternates depending on whether   $(i,j,k)$ contains two sites in sublattice A or two sites in sublattice B. Thus, the perturbation preserves the symmetry that combines time reversal with sublattice inversion, similarly to the effects of a staggered magnetic field  \cite{Takikawa2019,Nakazawa2022,yokoyama2025}.

 The analytical solution   is obtained by mapping the spin operators at each site onto four Majorana fermions $\left(b_i^x,b_i^y,b_i^z,c_i\right)$. The latter are Hermitian,  obey canonical anticommutation relations, and square to the identity: $ (b_{i}^{\alpha} )^{2}= (c_{i} )^{2}=1$. The components of the spin operators are represented by  \begin{equation}
  \sigma_i^\alpha= \ii b^\alpha_i c^{\phantom\dagger}_i,   
 \end{equation}
 along with the  local constraint $D_i=b^x_ib^y_ib^z_ic^{\phantom\dagger}_i=1$ $\forall i$, which selects   physical spin-1/2 states.  In this    representation, the  Hamiltonian $H'_K=H_K+H_{\rm pert}$ assumes the form 
\bea
H'_K &=&\ii\sum_\alpha J_\alpha  \sum_{\left\langle i,j\right\rangle_\alpha } \hat{u}^\alpha_{ij}c_{i}c_{j}\nonumber\\
&&-\ii\sum_{\alpha}J'_\alpha\sum_{\langle i,j\rangle_\alpha\langle j,k\rangle_\beta } (-1)^{s_i} \hat u^\alpha_{ij} \hat u_{jk}^\beta c_i c_k,
\label{eq:Hmajoranas}
\eea
where the operators $\hat{u}^\alpha_{ij}=\text{i}b^\alpha_ib^\alpha_j$ are associated with the  bonds $\langle i,j\rangle_\alpha$. They   commute with the Hamiltonian and have eigenvalues $u^\alpha_{ij}=\pm1$ . Here $\hat{u}^\alpha_{ij}$ are $\mathbb{Z}_2$ gauge fields  in the sense that they change sign  under a local transformation $ (c_{j},b_{j}^{\alpha} )\mapsto (-c_{j},-b_{j}^{\alpha} )$, but  physical observables that involve the spin operator  $\sigma^\alpha_j$  remain invariant. 

\begin{figure}[t]
\includegraphics[width=.95\columnwidth]{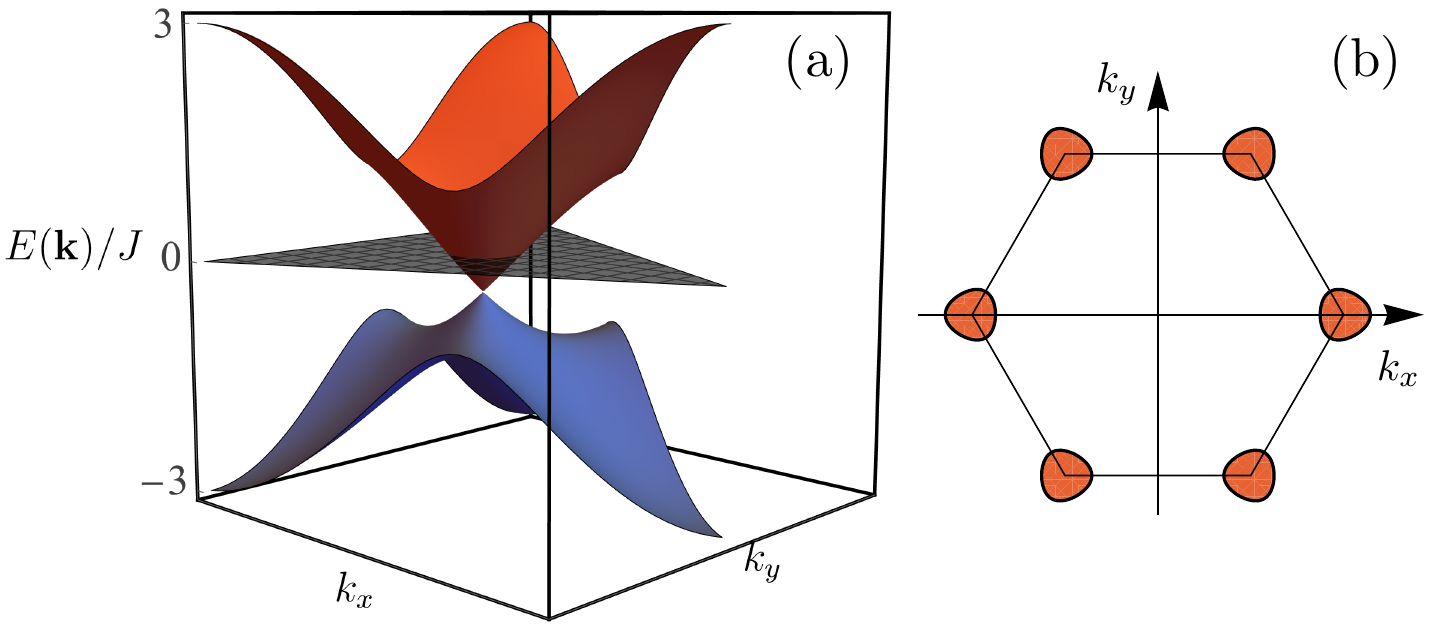}  
     \caption{(a) Dispersion of  Majorana fermions in the perturbed KHM with isotropic couplings   and $J'=0.1J$. The two bands are represented in one half of the Brillouin zone centered at the $K$ point, corresponding  to  momentum $\mb q_0=(\frac{2\pi}3,\frac{2\pi}{\sqrt3})$. (b) Majorana Fermi surface represented in the Brillouin zone.  }
    \label{f2}
\end{figure}

In the flux-free sector, we can choose a uniform gauge configuration $u^\alpha_{ij}=1$ for all nearest-neighbor bonds with $i\in \text{A}$ and $j\in\text{B}$. We then obtain a translation-invariant quadratic Hamiltonian for the $c$ Majorana fermions:\bea
H_{0}^{K}&=&\sum_\alpha \ii J_\alpha  \sum_{\substack{\left\langle i,j\right\rangle \\ i\in \text{A}}} c_{i}c_{j}- \sum_\alpha \ii J'_\alpha\sum_{ \llangle i,k\rrangle } (-1)^{s_i}  c_i c_k ,
\eea
where in   the   next-nearest-neighbor bonds $\llangle i,k\rrangle$  the vector $\mb r_{ik}$ pointing from site $i$ to site $k$ belongs to the set   $\mb r_{ik}\in \{(-1,0),(\frac12,-\frac{\sqrt{3}}2),(\frac12,\frac{\sqrt{3}}2)\}$.  Let us denote by   $c_{\nu}(\mb R)$  the Majorana fermion  associated with  sublattice $\nu=\text{A,B}$ in the unit cell at position  $\mb R$.
We  use  the Fourier mode expansion\bea
 c_{\nu} (\mb R)&=&\frac1{\sqrt{N}} \sum_{\mb k\in \text{BZ}} c^{\phantom\dagger}_{\nu,\mb k} e^{\ii\mb k\cdot \mb R},
 \eea
 where BZ stands for  Brillouin zone. Fermionic modes with opposite momenta are related by   $c^{\dagger}_{\nu,\mb k} =c^{\phantom\dagger}_{\nu,-\mb k} $. The Hamiltonian in the flux-free sector can be cast in  the form
\begin{equation}
H_{0}^{K}= \frac12\sum_{\mathbf{k}\in{\rm BZ}}\Psi_{\mathbf{k}}^{\dagger}h_K (\mathbf{k} )\Psi^{\phantom\dagger}_{\mathbf{k}},
\label{eq:HkKitaev}
\end{equation}
where $\Psi_\mathbf{k}=\left(c_{\text{A},\mathbf{k}},c_{\text{B},\mathbf{k}}\right)^T$ and \begin{equation}
h_K (\mathbf{k} )=\left(\begin{array}{cc}
\zeta(\mb k) & \gamma^{*} (\mathbf{k} )\\
\gamma (\mathbf{k} ) & \zeta(\mb k)
\end{array}\right).
\label{eq:hk}
\end{equation}
Here we define the functions  \bea
\gamma(\mathbf{k})&=&-\ii\left(J_{x}e^{\text{i}\mb k\cdot \mb a_1}+J_{y}e^{\text{i}\mb k\cdot \mb a_2}+J_{z}\right),\\
\zeta(\mb k)&=&2  J'_x \sin[\mb k\cdot (\mb a_2-\mb a_1)]-2  J'_y \sin(\mb k\cdot \mb{a}_2)\nonumber\\
&&+2  J'_z \sin(\mb k\cdot  \mb a_1),
\eea
where $\mb a_1$ and $\mb a_2$ are the lattice vectors shown in Fig. \ref{lattice}.  Diagonalizing the Hamiltonian, we obtain two bands with dispersion $E_{\pm}(\mb k)=\zeta(\mb k)\pm |\gamma(\mb k)|$.

Let us focus on the spatially isotropic case and set $J_\alpha=J>0$ and $J'_\alpha={J'}$ $\forall \alpha$. For $J'=0$, the spectrum is gapless with a Dirac cone at the  Brillouin zone vertex  $\mb q_0=(\frac{2\pi}3,\frac{2\pi}{\sqrt3})$. As we turn on $J'$, the Dirac point moves away from zero energy and one of the bands forms a Fermi surface centered at $\mb q_0$; see Fig \ref{f2}. Since the bands obey $E_+(-\mb k)=-E_-( \mb k)$, a copy of the Fermi surface appears centered at $-\mb q_0$. For $|J'|\ll J$, the Fermi surface is approximately circular with Fermi momentum $k_F \approx 6|J'|/J$. This behavior is analogous to the effect of a pseudoscalar potential in a modified Haldane model \cite{Colomes2018}.

In addition to itinerant  Majorana fermions, the KHM   has excitations associated with changing the flux configuration. The simplest flux excitation can be accomplished by flipping the eigenvalue of a single $\mathbb Z_2$ bond operator.   This procedure changes the flux in the two plaquettes that share the flipped bond and creates a pair of adjacent $\mathbb{Z}_2$ vortices,  \new{called visons \cite{KitaevAOP2006,takagi2019concept,Motome2020}.} The energy associated with this excitation defines the two-vison gap, probed, for instance,  in the dynamic spin structure factor  \cite{baskaran2007exact,knolle2014dynamics} and the conductance in scanning tunneling spectroscopy \cite{Feldmeier2020,Bauer2023}.  Importantly,  the presence of localized visons breaks translational symmetry. The Hamiltonian in the sector with two  visons adjacent to the $z$ bond in the $\mb R=\mb 0$ unit cell can be written as \be
H^{K}_{\rm 2v}=H_0^{K}+\hat V, \label{perturbHK}
\ee 
where 
\bea
\hat V&=&-2\ii J_z c_{\rm A}(\mb 0)c_{\rm B}(\mb 0)+2\ii J_z' c_{\rm A}(\mb 0)[c_{\rm A}(\mb a_1)-c_{\rm A}(\mb a_2)]\nonumber\\
&&-2\ii J_z' c_{\rm B}(\mb 0)[c_{\rm B}(-\mb a_1)-c_{\rm B}(-\mb a_2)]\label{scatterK}
\eea
can be viewed as a scattering potential for   itinerant Majorana fermions. Note that for $J_z'=0$ the   potential only involves Majorana fermions contained within the same unit cell; this   feature   enables the analytic calculation of the \new{two-vison gap} \cite{panigrahi2023analytic}, as we will discuss in Sec. \ref{sec:visongap}.

\subsection{\new{Chua-Yao-Fiete} model on the kagome lattice\label{sec:CYF}}

The  CYF model \cite{chua2011exact} has spins $3/2$ living on a kagome lattice. The interaction   depends on whether the  bond between    nearest-neighbor spins belongs to an up-pointing or down-pointing  triangle ($\Delta$ or $\nabla$, respectively). The Hamiltonian is given by  
\begin{eqnarray}
H_{\rm CYF}&=&J_{5}\sum_{i}\Gamma_{i}^{5}+J_{\Delta}\sum_{\left\langle i,j\right\rangle \in\Delta}\Gamma_{i}^{1}\Gamma_{j}^{2}+J_{\nabla}\sum_{\left\langle i,j\right\rangle \in\nabla}\Gamma_{i}^{3}\Gamma_{j}^{4}
\nonumber\\
&&+J_{\Delta}^{\prime}\sum_{\left\langle i,j\right\rangle \in\Delta}\Gamma_{i}^{15}\Gamma_{j}^{25}+J_{\nabla}^{\prime}\sum_{\left\langle i,j\right\rangle \in\nabla}\Gamma_{i}^{35}\Gamma_{j}^{45},
\end{eqnarray}
where $\Gamma^\mu$ with $\mu=1,\dots,5$ are $\Gamma$ matrices obeying  the Clifford algebra, $ \{ \Gamma^{\mu},\Gamma^{\mu'} \} =2\delta^{\mu\mu'}$, and $\Gamma^{\mu\mu'}=\frac1{2\ii}[\Gamma^{\mu},\Gamma^{\mu'}]$. They   are related to  spin-$3/2$ operators by
\bea
\Gamma^{1}&=&\frac{1}{\sqrt{3}}\left\{ S^{y},S^{z}\right\},  \quad  \Gamma^{2}=\frac{1}{\sqrt{3}}\left\{ S^{z},S^{x}\right\}, \nonumber\\
\Gamma^{3}&=&\frac{1}{\sqrt{3}}\left\{ S^{x},S^{y}\right\} , \quad  \Gamma^{4}=\frac{1}{\sqrt{3}}\left[\left(S^{x}\right)^{2}-\left(S^{y}\right)^{2}\right],\nonumber\\
\Gamma^{5}&=&\left(S^{z}\right)^{2}-\frac{5}{4}.   
\eea

\begin{figure}[t]
    \centering
    \subfigure[]{\includegraphics[scale=0.25]{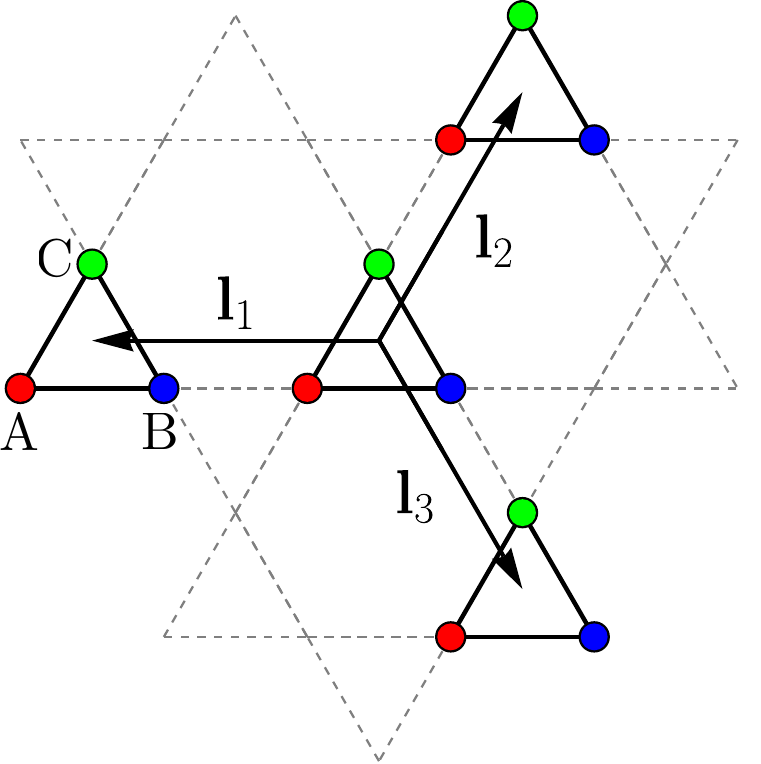} \label{fig:kagome1}}
    \hspace{0.3cm}
 \subfigure[]{\includegraphics[scale=0.27]{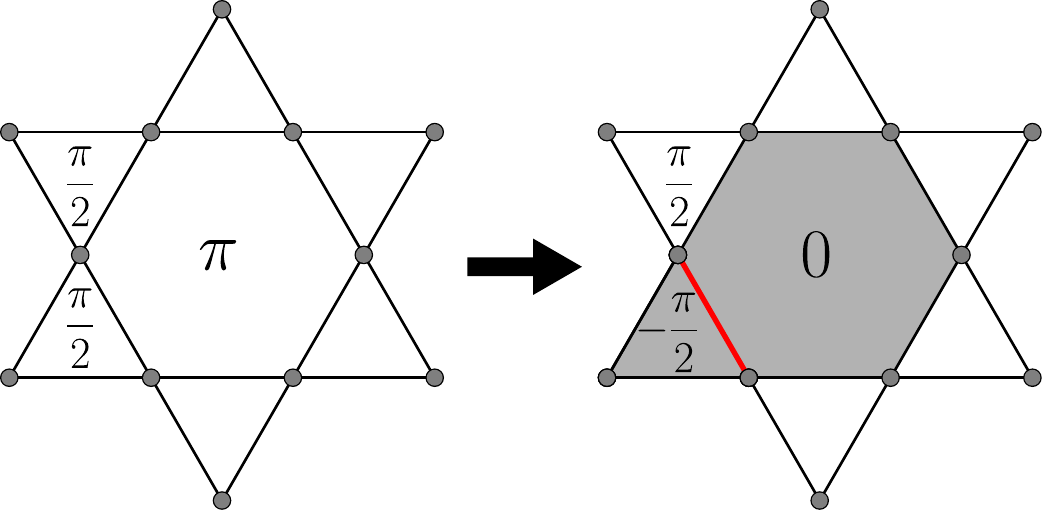}\label{fig:kagome2}}
    \caption{Kagome lattice. (a) The three sublattices, A, B, and C, are represented in red, blue, and green, respectively. In terms of $\mb a_1=(\frac12,\frac{\sqrt3}2)$ and $\mb a_2=(-\frac12,\frac{\sqrt3}2)$, we define the vectors  $\mathbf{l}_1=\mathbf{a}_2-\mathbf{a}_1$,  $\mathbf{l}_2=\mathbf{a}_1$, and $\mathbf{l}_3=-\mathbf{a}_2$. (b) Creation of a $\left(\Delta,\hexagon\right)$ vison pair on the $\left\{ \frac{\pi}{2},\frac{\pi}{2},\pi\right\} $ flux configuration.}
    \label{fig:kagome}
\end{figure}

The CYF model is exactly solvable by the same strategies developed in the KHM. We can identify three sets of  plaquette operators that commute with the Hamiltonian:
\bea
W_{\Delta}&=&\Gamma_{i}^{12}\Gamma_{j}^{12}\Gamma_{k}^{12},\\
W_{\nabla}&=&\Gamma_{i}^{34}\Gamma_{j}^{34}\Gamma_{k}^{34},\\
W_{\hexagon}&=&\Gamma_{i}^{23}\Gamma_{j}^{14}\Gamma_{k}^{23}\Gamma_{l}^{14}\Gamma_{m}^{23}\Gamma_{n}^{14},
\eea
where the sites in the hexagonal plaquettes are ordered as belonging to sublattices ABCABC; see Fig. \ref{fig:kagome1}. 
The spin operators can be represented in terms of six Majorana fermions $(\xi_{i}^{1},\xi_{i}^{2},\xi_{i}^{3},\xi_{i}^{4},c_{i},d_{i})$ as\bea
\Gamma^a_{i}=\ii \xi^a_ic^{\phantom a}_i, \quad \Gamma^5_{i}=\ii c^{\phantom a}_id^{\phantom a}_i,\quad  \Gamma^{a5}_{i}=\ii \xi^{a}_id^{\phantom a}_i, 
\eea
for $a=1,\dots,4$. Similarly to the KHM,  we remove  the unphysical extension of the Hilbert space by imposing a constraint   $G_i=-\ii\xi_i^{1}\xi_i^{2}\xi_i^{3}\xi_i^{4}c^{\phantom a}_id^{\phantom a}_i=1$ for all sites. In this representation, the Hamiltonian reads 
\begin{equation}
H_{\rm CYF}=\sum_{\left\langle i,j\right\rangle }\left(\ii J_{ij}\hat{v}_{ij}c_{i}c_{j}+\ii J_{ij}^{\prime}\hat{v}_{ij}d_{i}d_{j}\right)+\ii J_{5}\sum_{i}c_{i}d_{i},
\label{eq:Hfermionslivres}
\end{equation}
with $J_{ij}=J_{\nabla},J_{\Delta}$ and   $J_{ij}^{\prime}=J_{\nabla}^{\prime},J_{\Delta}^{\prime}$ 
depending on the the bond $\langle i,j\rangle$. The   $\mathbb{Z}_2$ gauge fields $\hat{v}_{ij}$  associated with  the kagome lattice bonds are  defined by 
\begin{equation}
\hat{v}_{ij}=\begin{cases}
\ii\xi_{i}^{1}\xi_{j}^{2}, & \text{if }\left\langle i,j\right\rangle \in\Delta,\\
\ii\xi_{i}^{3}\xi_{j}^{4}, & \text{if }\left\langle i,j\right\rangle \in\nabla.
\end{cases}
\end{equation}

The flux configuration in the ground state sector depends on the interaction parameters.  If we write the eigenvalues of the plaquette operators as $w_p=e^{\ii\varphi_{p}}$,   the possible values of the fluxes  are
\begin{equation}
\varphi_{p}=\begin{cases}
\pm\frac{\pi}{2}, & p\in\left\{ \Delta,\nabla\right\} ,\\
0,\pi, & p=\hexagon.
\end{cases}
\end{equation}
Note that the fluxes through triangular plaquettes spontaneously break time reversal symmetry, and a configuration $\left\{ \varphi_{\Delta},\varphi_{\nabla},\varphi_{\hexagon}\right\} $ is degenerate with  its time reversal conjugate $\left\{ -\varphi_{\Delta},-\varphi_{\nabla},\varphi_{\hexagon}\right\} $. Thus, there are four possible translation-invariant, nondegenerate   flux configurations: $
\left\{ \frac{\pi}{2},\frac{\pi}{2},\pi\right\},   \left\{ -\frac{\pi}{2},\frac{\pi}{2},\pi\right\}, 
\left\{ \frac{\pi}{2},\frac{\pi}{2},0\right\},   \left\{ -\frac{\pi}{2},\frac{\pi}{2},0\right\}$. 

We   focus on a parameter regime where  the $\left\{ \frac{\pi}{2},\frac{\pi}{2},\pi\right\} $ configuration, shown in Fig. \ref{fig:kagome2},  gives the ground state.  In this case, we can fix the eigenvalues   $  v_{ij}=\pm1$ in a gauge configuration that preserves the translational symmetry of the kagome lattice. Specifically, we set $v_{ij}=1$ whenever $i\in \text{A}$ or $j\in\text{C}$.  Equation  (\ref{eq:Hfermionslivres}) then yields  a quadratic Hamiltonian  for $c$ and $d$ Majorana fermions, which we denote by $H_0^{\rm CYF}$. Taking a Fourier transform, we obtain a  $6\times6$ Bloch matrix: 
\begin{equation}
h_{\text{CYF}} (\mathbf{k} )=\left(\begin{array}{cc}
\mathcal{H}_{cc}(\mb k) & \mathcal{H}_{cd}\\
\mathcal{H}^{\dagger}_{cd} & \mathcal{H}_{dd}(\mb k)
\end{array}\right),
\label{eq:tildeh}
\end{equation}
for $\mb k\in \text{HBZ}$.  Here we define the matrices
\bea
\mathcal{H}_{cc}(\mb k)&=&\left(\begin{array}{ccc}
0 & f_1(\mb k) & f^*_2(\mb k)\\
f_1^{*}(\mb k) & 0 & f_3(\mb k)\\
f_2(\mb k) & f_3^{*}(\mb k) & 0
\end{array}\right),\\
\mathcal{H}_{cd}&=&\ii\left(\begin{array}{ccc}
J_{5} & 0 & 0\\
0 & J_{5} & 0\\
0 & 0 & J_{5}
\end{array}\right).
\eea
The form factor functions in $\mathcal{H}_{cc}(\mb k)$ are given by  $f_n (\mb k )=\ii (J_{\Delta}+J_{\nabla}e^{\ii\mathbf{k}\cdot\mathbf{l}_{n}})$, with the vectors $\mb l_n$ represented in Fig. \ref{fig:kagome1}. 
 The matrix  $\mathcal{H}_{dd}(\mb k)$ is obtained from  $\mathcal{H}_{cc}(\mb k)$ by taking  $J_{\Delta,\nabla}\to J_{\Delta,\nabla}^\prime$. Diagonalizing $h_{\text{CYF}}\left(\mathbf{k}\right)$, we obtain six bands with dispersion relations $E_\lambda (\mathbf{k} )$, with $\lambda=1,\dots,6$ in increasing order of energy.

We can find regimes in which a low-energy band   crosses zero energy by tuning the coupling constants.  In the following,  we  fix $J_\nabla=J^\prime_\nabla=1$,  $J_\Delta=J^\prime_\Delta=0.5$ and tune the parameter $J_5$. In this case, we find that  for $J_5=\bar J_5\approx 0.865$ the two lowest-energy bands touch at a Dirac point at zero energy, similarly to the spectrum of the KHM.  We modulate the size of the Majorana Fermi surface by varying  $J_5$ around this value. For $|J_5-\bar J_5|\ll J_{\Delta}$, the Fermi surface is approximately circular with radius $k_F\propto |J_5-\bar J_5|$.

To study visons in the CYF model, we have to account for  the   three   families of plaquettes. Here we   choose a pair of visons   created by flipping the sign of $v_{ij}$ at the bond shared by of adjacent $\left(\Delta,\hexagon\right)$ plaquettes, as shown in Fig. \ref{fig:kagome2}. Similarly to Eq. (\ref{scatterK}), we   write the Hamiltonian in this two-vison sector as\be
H_{\rm 2v}^{\rm CYF}=H_{0}^{\rm CYF}-2\ii J_{\Delta}c_{\rm B}(\mb 0)c_{\rm C}(\mb 0)-2\ii J'_{\Delta}d_{\rm B}(\mb 0)d_{\rm C}(\mb 0),\label{hcyf}
\ee
which involves only Majorana fermions in sublattices B and C within the $\mb R=0$ unit cell. Similar results can be obtained for a  $\left(\nabla,\hexagon\right)$ vison pair     by changing the choice of the unit cell.

\section{Two-vison gap\label{sec:visongap}}

In this section, we analyze the two-vison gap in the KHM and CYF models using   numerical diagonalization for finite-size lattices and an analytical approach based on scattering phase shifts.  We compare the results of the two approaches and analyze the influence of a Majorana Fermi surface on   the \new{two-vison gap}. 

\subsection{Visons in the Kitaev honeycomb model \label{pureKitaev}}

\begin{figure}[t]
    \centering
    \subfigure[]{\includegraphics[width=.98\columnwidth]{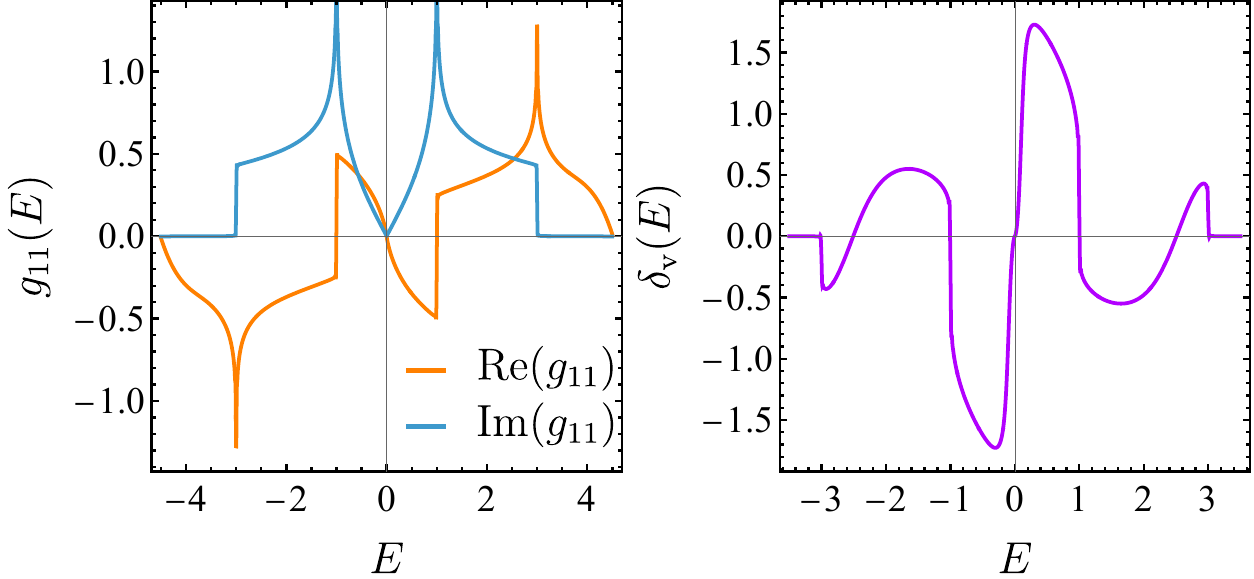}} 
    \subfigure[]{\includegraphics[width=.98\columnwidth]{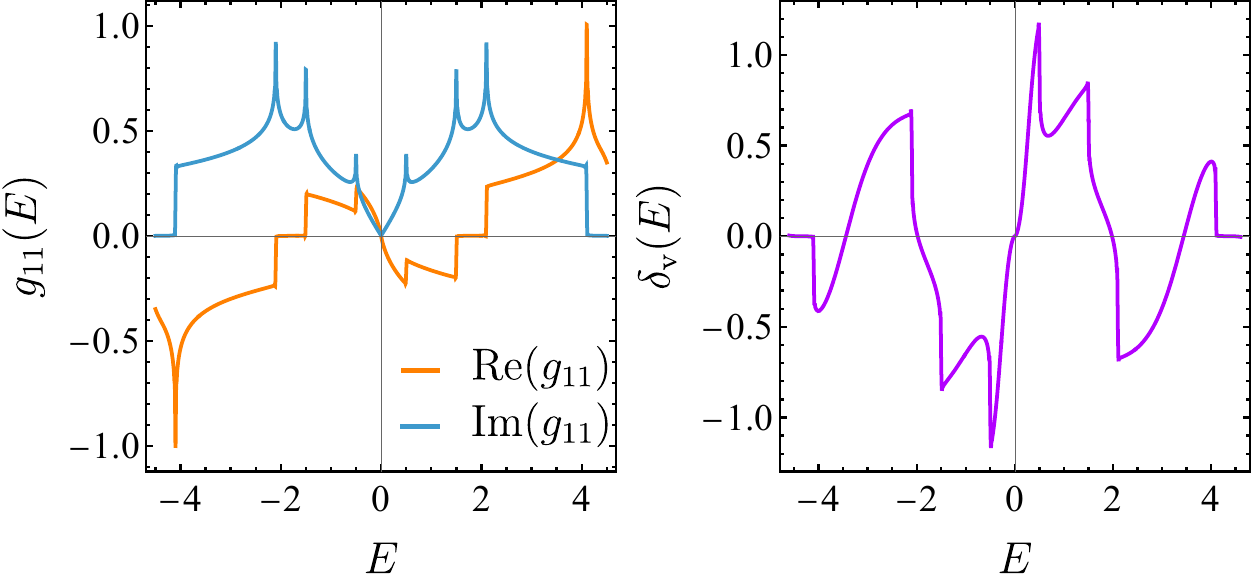}} 
    \caption{Matrix element $g_{11}(E)$ of the local Green's functions and phase shift $\delta_\text{v} (E )$ in the unperturbed KHM ($J_\alpha'=0$) for two sets of coupling constants: (a) isotropic model with $J_x=J_y=J_z=1$; (b) anisotropic model in the gapless phase with $J_x=1.8$, $J_y=1.3$, and $J_z=1$.}
    \label{fig:phashkitaev}
\end{figure}

Let us briefly review the analytical approach for the \new{two-vison gap} in the unperturbed KHM  \cite{panigrahi2023analytic}, obtained by setting $J_\alpha'=0$ in Eqs. (\ref{perturbHK}) and  (\ref{scatterK}).  The key idea is to relate the \new{two-vison gap} to the phase shift for Majorana fermions scattering off the local  potential   in the Hamiltonian for the two-vison sector. For $J_\alpha'=0$, we rewrite the Hamiltonian  as\begin{equation}
    H_{2\text{v}}=\frac{1}{2}\sum_{\mathbf{k}}\Psi_{\mathbf{k}}^{\dagger}h_K(\mathbf{k} )\Psi^{\phantom\dagger}_{\mathbf{k}}+\frac{1}{2} {c}^{T}(\mb 0)\left(2J_{z}\tau_{y}\right) {c}(\mb 0),
\end{equation}
where $ {c}(\mb 0)=\left(c_{A}(\mb 0),c_{B}(\mb 0)\right)^T$ and $\tau_y$ is the Pauli matrix acting  in sublattice space. We introduce the noninteracting Green's function \be
G_0(\mb k,z)=[ z\mathbb I_2-h_{K}(\mb k)]^{-1},
\ee
where $z$ is a complex frequency and $\mathbb I_2$ is the $2\times2$ identity matrix.  Notice that the  scattering  potential  is completely local in real space, which implies the absence of momentum dependence in reciprocal space.  In this case, one can analytically solve the Dyson equation for the exact  Green's function, as done in the context of impurity problems  \cite{dewitt1956transition,doniach1998green} . The solution is  expressed  in terms of the local   Green's function, defined as  \be
g(z)=\frac1{N}\sum_{\mb k}G_0(\mb k,z).
\ee
In the thermodynamic limit, it is  convenient to  compute the integral over the Brillouin zone as   
\begin{equation}
    g (z)=\frac{1}{4\pi^2}\int_{0}^{2\pi}dk_{1}\int_{0}^{2\pi}dk_{2}\, G_0(\mb k,z),
    \label{gnunu}
\end{equation}
where $k_1=\mb k\cdot \mb a_1$ and  $k_2=\mb k\cdot \mb a_2$. 
 The phase shift is   given by \cite{panigrahi2023analytic} 
\begin{equation}
    \delta_{\text{v}} (E )=\text{Im}\left\{ \ln\left[\det\left( \mathbb{I}_2-\hat{V}_0\,g (E+i\eta )\right)\right]\right\},\label{phaseshift}
\end{equation}
where  $\hat{V}_0=2J_z\tau_y$  and we take $\eta\to0^+$. One can then use Fumi's theorem \cite{fumi1955cxvi,mahan2013many,maslov2022impurity,liu2020theory} to calculate the change in the  free energy due to the presence of the scattering potential by integrating  the phase shift. At zero temperature, the result    is exactly  the gap $\Delta_{\rm v}$ for two adjacent visons in the thermodynamic limit, given by \bea
\Delta_{\rm v}&=& \frac1{2\pi}\int_{-\infty}^{\infty}dE \, \delta_{\rm v}(E)\text{ sgn}(E) .\label{phaseshiftKHM}
\eea

\begin{figure}[t]
    \begin{center}
   \includegraphics[width=.95\columnwidth]{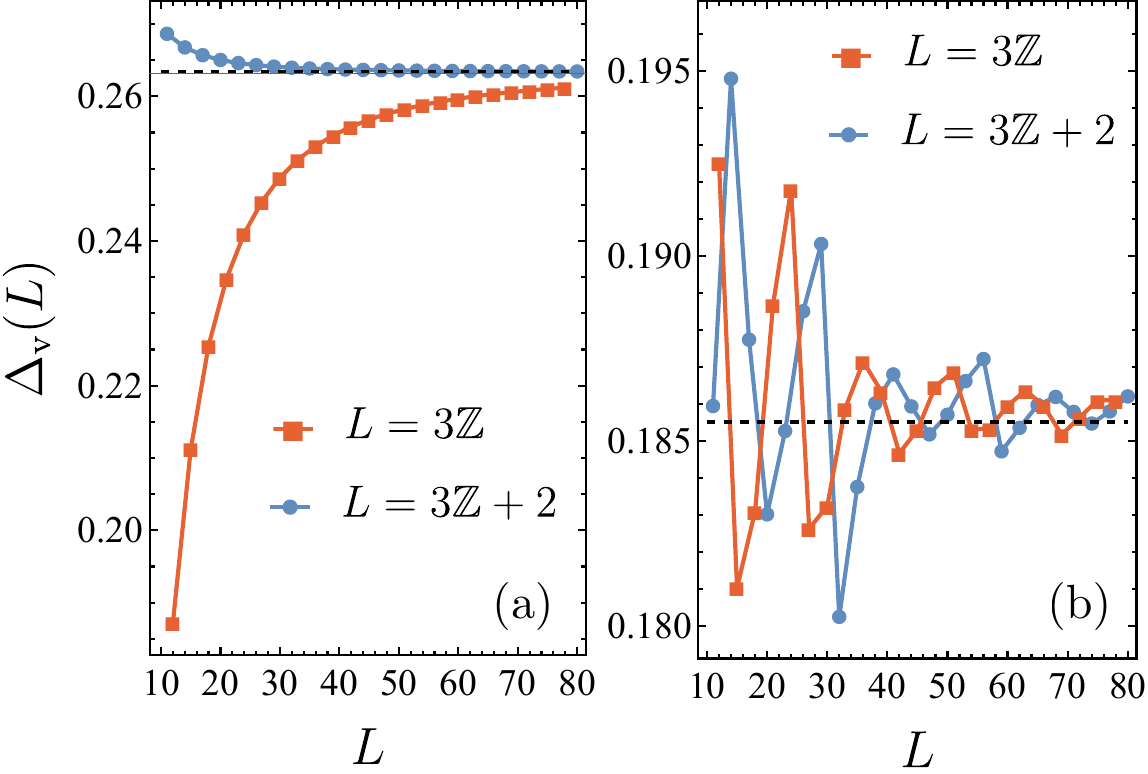}
    \caption{\new{Two-vison gap} in the unperturbed KHM ($J_\alpha'=0$).  The symbols represent   exact diagonalization results obtained for  lattices with $N=L^2$ unit cells. We consider two   sets of coupling constants: (a) the isotropic case $J_x=J_y=J_z=1$; the dashed line represents the   result $\Delta_{\rm v}=0.2633$   from the analytical approach in Ref. \cite{panigrahi2023analytic}; (b) an anisotropic case in the gapless phase $J_x=1.8$, $J_y=1.3$, $J_z=1$; the dashed line represents $\Delta_{\rm v}=0.1855$, which we obtained using Eq. (\ref{phaseshiftKHM}).\label{fig:Diracgap}}
    \end{center}
\end{figure}

Figure \ref{fig:phashkitaev}  illustrates the energy dependence of the local Green's function and the phase shift in the unperturbed Kitaev model. \new{Here we set the broadening parameter to $\eta=10^{-3}$.} In addition to the isotropic model $J_x=J_y=J_z$ studied in Ref. \cite{panigrahi2023analytic},    we consider an example of anisotropic coupling constants in the regime $0<J_z<J_y<J_x$. When the coupling constants obey the triangle inequality   $J_x<J_y+J_z$, the system remains in the gapless phase   \cite{KitaevAOP2006}, but the Dirac cone in the Majorana dispersion is located  at an incommensurate wave vector.   As shown in  Fig. \ref{fig:phashkitaev}, we find  $\delta_{\rm v}(-E)=-\delta_{\rm v}(E)$ as a consequence of particle-hole symmetry, and  the phase shift is nonzero for   $|E|<E_{\rm max}$, where $E_{\rm max}$  is the maximum energy in the Majorana fermion bands.

In Fig. \ref{fig:Diracgap}, we compare the  \new{two-vison gap} from Eq. (\ref{phaseshiftKHM}) with the result obtained by diagonalizing the Hamiltonian on finite lattices. We perform the   integrals in Eqs. (\ref{gnunu}) and (\ref{phaseshiftKHM}) numerically using  the Gauss-Kronrod method, for values of $|z|<E_{\max}$ with step $0.01$ interpolated with a cubic spline. In the  purely numerical method, we consider finite lattices with periodic boundary conditions and $L$ unit cells in the directions of   lattice vectors $\mb a_1$ and $\mb a_2$ and compute the  gap as function of system size,   $\Delta_{\rm v}(L)$. We have verified that in the strongly anisotropic regime $J_x>J_y+J_z$, where the  KHM is in a gapped Abelian phase \cite{KitaevAOP2006}, $\Delta_{\rm v}(L)$ is a smooth function   that converges exponentially fast to its asymptotic value. For the gapless isotropic point, we separate the data with $L \text{ mod }3=0$ from $L \text{ mod }3\neq0$  because in the former   the fermionic spectrum contains a  state with momentum   $\mb q_0$, which has exactly zero energy and affects the finite size scaling. 
In this case,   the convergence  of the \new{two-vison gap} is slower when $L \text{ mod }3=0$. Nevertheless, both curves for $L \text{ mod }3=0$ and $L \text{ mod }3=2 $ are  smooth and can be reliably extrapolated to the limit $L\to\infty$;   the result agrees with the estimate $\Delta_{\rm v}=0.2633$ from Ref. \cite{panigrahi2023analytic}.  By contrast, for  the anisotropic model  with $J_x<J_y+J_z$ we observe oscillations in $\Delta_{\rm v}(L)$ with a frequency that depends on the incommensurate wave vector at the Dirac point.  While the extrapolation of the finite-size data is not so clear in this case,  the result  is consistent with the calculation of $\Delta_{\rm v}$ from Eq. (\ref{phaseshiftKHM})   within our numerical precision.

\begin{figure}[t]
    \begin{center}
\includegraphics[width=.85\columnwidth]{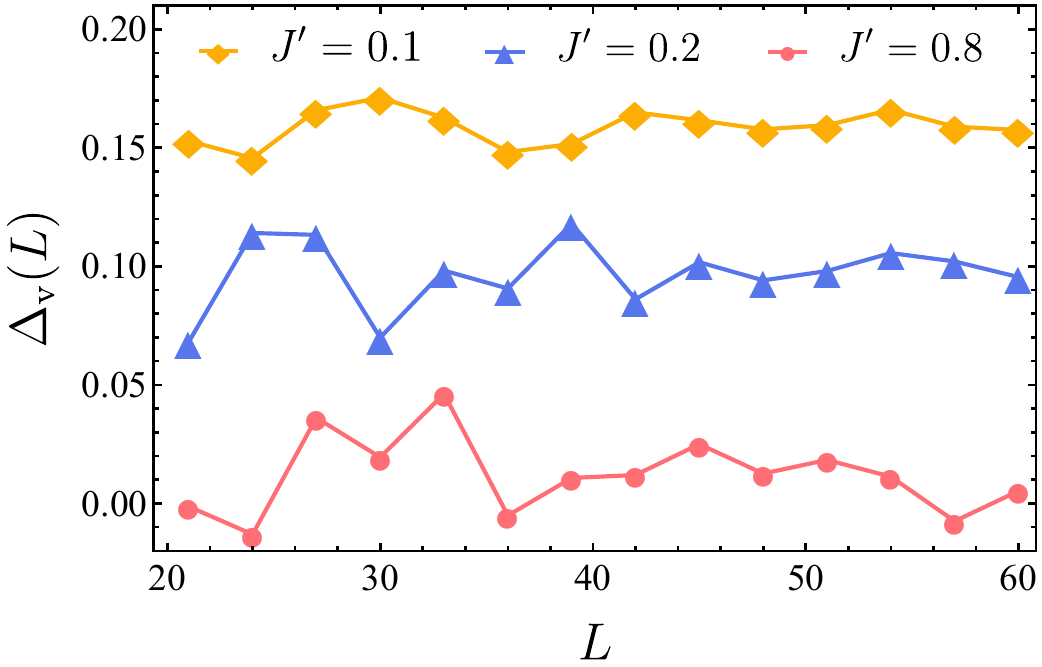}
    \caption{\new{Two-vison gap} in the perturbed KHM with isotropic couplings $J_\alpha=1$ and $J'_\alpha=J'$, for different values of $J'$.  Here we select system sizes with $L$ mod $3=0$.    \label{fig:pertKHMgap}}
    \end{center}
\end{figure}

We now turn to the  calculation of the  \new{two-vison gap} for the perturbed KHM, which allows us to explore the effect of the Majorana Fermi surface.  We focus on the isotropic model, setting  $J_\alpha=1$ and $J'_\alpha=J'>0$. \new{In this case, the analytical approach described in Sec. \ref{pureKitaev} is not directly applicable because the scattering potential couples Majorana modes belonging to different unit cells. Since the potential remains finite-ranged, the method can, in principle, be extended by enlarging the unit cell to include all  Majorana modes that appear in the impurity term.  However, in the present work we restrict our analysis to calculating the two-vison gap on finite lattices.} The result for different values of $J'$ is shown in Fig. \ref{fig:pertKHMgap}. To interpret the result, we recall that the size of the Majorana Fermi surface increases with $J'$. We see that the finite-size effects become more severe with  increasing $J'$ as the gap exhibits slowly decaying oscillations. Oscillating correlations  are expected in QSLs with  Majorana Fermi surfaces \cite{lai2011power}, and the effect on the \new{two-vison gap} observed here  is   reminiscent of oscillations in  the   Casimir effect for fermionic systems \cite{Beenakker2024,nakayama2023dirac}.

Despite the oscillations, it is clear in Fig. \ref{fig:pertKHMgap} that the asymptotic value of the \new{two-vison gap} decreases with $J'$, and it appears to vanish for $J'\approx 0.8J$. In this exactly solvable model, where we have access to  the eigenstates in all flux sectors,  we expect the vanishing gap  to signal a proliferation of visons, implying a global change in the flux configuration of the ground state. To test this idea, we calculate the energy density for a $\pi$-flux state corresponding to the lowest-energy state in the sector with   $\varphi_p=\pi$  $\forall p$. The latter can be described by a gauge configuration with a four-site unit cell in which we flip the sign of $u_{ij}^z$ on every other $z$ bond.  Indeed, we find that the energy density of the $\pi$-flux state becomes lower than that of the flux-free state for $J'\gtrsim 0.69J$.  Thus, the system undergoes a first-order phase transition before the gap for local two-vison excitations closes. Interestingly, the $\pi$-flux QSL   state for  $J'\gtrsim 0.69J$ is  also gapless, but has  a significantly smaller Majorana Fermi surface than the flux-free state below the transition. \new{While this analysis shows that the zero-flux sector becomes unstable for sufficiently large $J'$, we cannot rule out the possibility of phase transitions into  more complex flux configurations, such as vison crystals \cite{Zhang2019}, in some parameter regimes.}

\subsection{ Visons in the \new{Chua-Yao-Fiete} model}

To further  investigate  a relation between the \new{two-vison gap} and the size of the Majorana Fermi surface, we now consider the CYF model. In this case, the scattering potential in the two-vison sector, see Eq. (\ref{hcyf}), is completely local in real space. We can then straightforwardly generalize the analytical approach of Ref. \cite{panigrahi2023analytic} to this multiband system. The local Green's function   becomes\be
g_{\rm CYF}(z)=\frac1{4\pi^2}\int_0^{2\pi}dk_1\int_{0}^{2\pi}dk_2\,[z\mathbb I_6-h_{\rm CYF}(\mb k)]^{-1},
\ee
where $\mathbb I_6$ is the $6\times6$ identity matrix. The phase shift is now given by 
\be
\delta_{\rm v}(E)= \text{Im}\left\{ \ln\left[\det\left( \mathbb{I}_6-\hat{\mathcal{V}}g_{\text{CYF}}\left(E+i\eta\right)\right)\right]\right\},
\ee
where the matrix that encodes the creation of a $\left(\Delta,\hexagon\right)$ vison pair is 
\begin{equation}
    \mathcal{V}=\left(\begin{array}{cccccc}
0 & 0 & 0 & 0 & 0 & 0\\
0 & 0 & -2iJ_{\Delta} & 0 & 0 & 0\\
0 & 2iJ_{\Delta} & 0 & 0 & 0 & 0\\
0 & 0 & 0 & 0 & 0 & 0\\
0 & 0 & 0 & 0 & 0 & -2iJ_{\Delta}^{\prime}\\
0 & 0 & 0 & 0 & 2iJ_{\Delta}^{\prime} & 0
\end{array}\right).
\label{eq:mathcalV}
\end{equation}
As in the KHM,  the phase shift in the CYF model obeys $\delta_{\rm v}(-E)=-\delta_{\rm v}(E)$ and is nonzero only within the energy bands of the Majorana fermions.  After calculating the phase shift, we compute the \new{two-vison gap} using  the integral in Eq. (\ref{phaseshiftKHM}). \new{As done for the Kitaev model, we use a Gauss-Kronrod method with integration step 0.01 and broadening parameter $\eta=10^{-3}$.}

\begin{figure}
    \centering
  \includegraphics[width=.85\columnwidth]{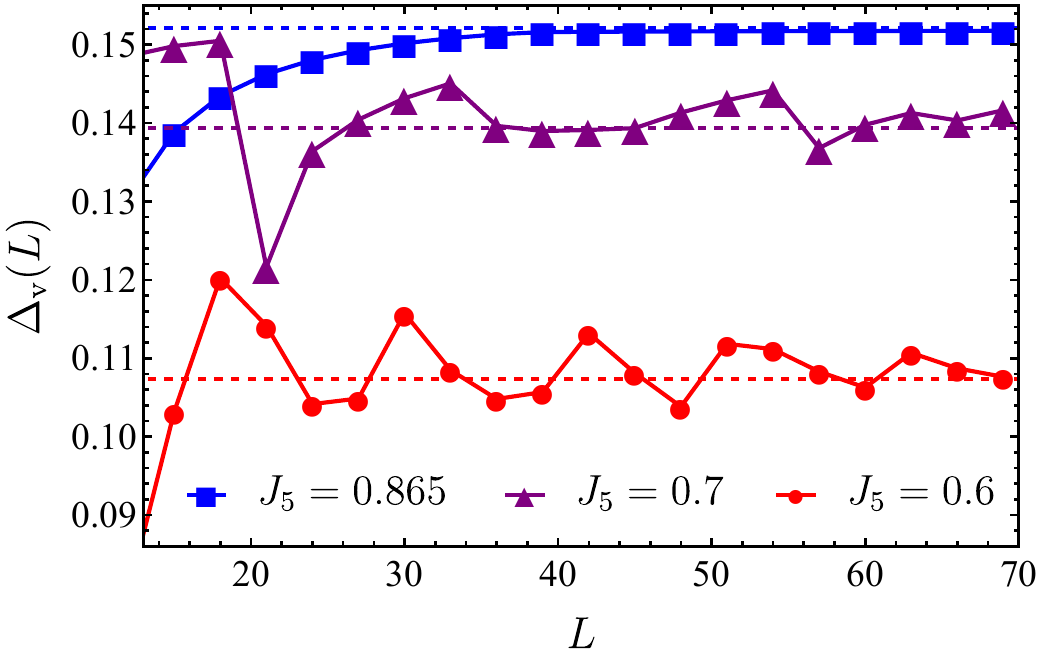}
    \caption{Gap for creating a  $\left(\Delta,\hexagon\right)$ vison pair  in the CYF model. Here we      fix $J_\nabla=J^\prime_\nabla=1$, $J_\Delta=J^\prime_\Delta=0.5$, and consider three   values of $J_5$.  The symbols represent the data  from exact diagonalization in lattices with  $L\text{ mod 3}=0$. The dashed lines represent the respective results from the analytical expression for $\Delta_{\rm v}$ in the thermodynamic limit. } 
    \label{fig:deltavnCYF}
\end{figure}

We use the parameters discussed in Sec. \ref{sec:CYF},  setting $J_\nabla=J^\prime_\nabla=1$ and $J_\Delta=J^\prime_\Delta=0.5$. The useful starting point for comparison with the KHM is $J_5=\bar J_5=0.865$, at which  the Dirac point lies at   zero energy. Note that $J_5$ does not appear in the scattering potential in Eq. (\ref{eq:mathcalV}); the main effect of varying $J_5$ is to control the size of the Majorana Fermi surface.   Figure \ref{fig:deltavnCYF} shows our results for the \new{two-vison gap} for three different  values of $J_5$. Similarly to the KHM, if we select system sizes with $L\text{ mod }3=0$, we observe a smooth dependence on $L$ when we tune $J_5=\bar J_5$, such that $k_F=0$. For other values of  $J_5$, the numerical method is, as expected, less conclusive because the \new{two-vison gap}  exhibits oscillations.  However, the  value around which  the gap oscillates   is consistent with the result from the analytical expression for $\Delta_{\rm v}$ in the thermodynamic limit; see the dashed lines in Fig.   \ref{fig:deltavnCYF}.

\begin{figure}[t]
    \centering
    \subfigure[]{\includegraphics[width=0.235\textwidth]{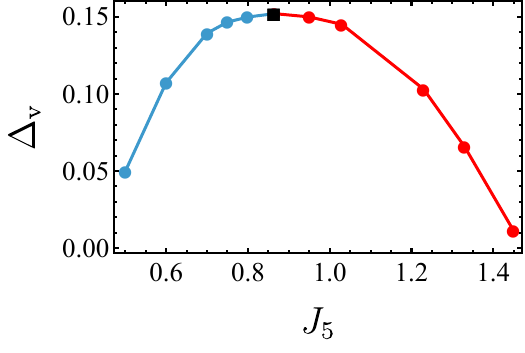}\label{figgapJ5}} 
    \subfigure[]{\includegraphics[width=0.235\textwidth]{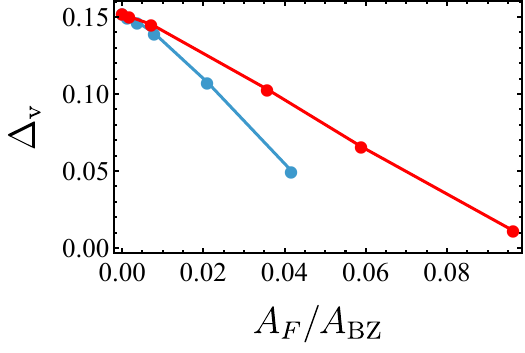}\label{kfvsj5}}
    \caption{\new{Two-vison gap} in the thermodynamic limit for the CYF model with $J_\nabla=J^\prime_\nabla=1$ and $J_\Delta=J^\prime_\Delta=0.5$.  (a) Gap as function of $J_5$. The black square corresponds to the point $J_5=0.865$, where the  Majorana Fermi surface shrinks to a Dirac point. Blue and red dots represent values of $J_5$ below and above this point, respectively. (b) \new{Two-vison gap} as a function of the \new{area $A_F$ enclosed by the Majorana Fermi surface, normalized by the Brillouin zone area $A_{\rm BZ}=8\pi^2/\sqrt{3}$.} }
    \label{fig:colored}
\end{figure}

In Fig. \ref{figgapJ5}, we show the \new{two-vison gap} calculated in the thermodynamic limit as a function of $J_5$. Remarkably, we find that  the \new{two-vison gap} is maximum at $J_5=\bar J_5$. Moving away from this point, we  plot the \new{two-vison gap} as a function of the area enclosed by the Fermi surface in Fig. \ref{kfvsj5}. \new{Here $A_F$ refers to the area enclosed by all symmetry-related pockets, and the axis is rescaled by the total Brillouin zone area $A_{\rm BZ}=8\pi^2/\sqrt{3}$.} Clearly,  the \new{two-vison gap} decreases as the Fermi surface grows, in agreement with  the     behavior     observed in the perturbed KHM. For sufficiently large Majorana Fermi surfaces, the gap approaches zero, which indicates that the  $\left\{\frac{\pi}2,\frac{\pi}2,\pi\right\}$ flux configuration must become unstable. Calculating the ground state energy density for all inequivalent flux configurations with fixed  $J_\nabla=J^\prime_\nabla=1$ and $J_\Delta=J^\prime_\Delta=0.5$, we find that   the regime where the $\left\{\frac{\pi}2,\frac{\pi}2,\pi\right\}$ configuration  corresponds to the ground state is bounded by $0.46<J_5<1.49$.  These phase boundaries are remarkably close to the points where the gap for local two-vison excitations  vanishes, which we estimate as $J_5\approx 0.44$ and $J_5\approx 1.48$. 

\section{Spin correlations\label{sec:correl}}

The results presented in Sec. \ref{sec:visongap} indicate that the \new{two-vison gap} decreases as the Majorana Fermi surface increases. Importantly,  the gap can be expressed as an integral of the exact phase shift, which sums up Majorana-vison scattering processes to all orders. This relation suggests that the effective scattering amplitude becomes weaker for larger Majorana Fermi surfaces. To probe this effect, in this section we  consider local spin correlations in the vicinity of a vortex pair in the perturbed KHM. 

\begin{figure}
    \centering
  \includegraphics[width=.85\columnwidth]{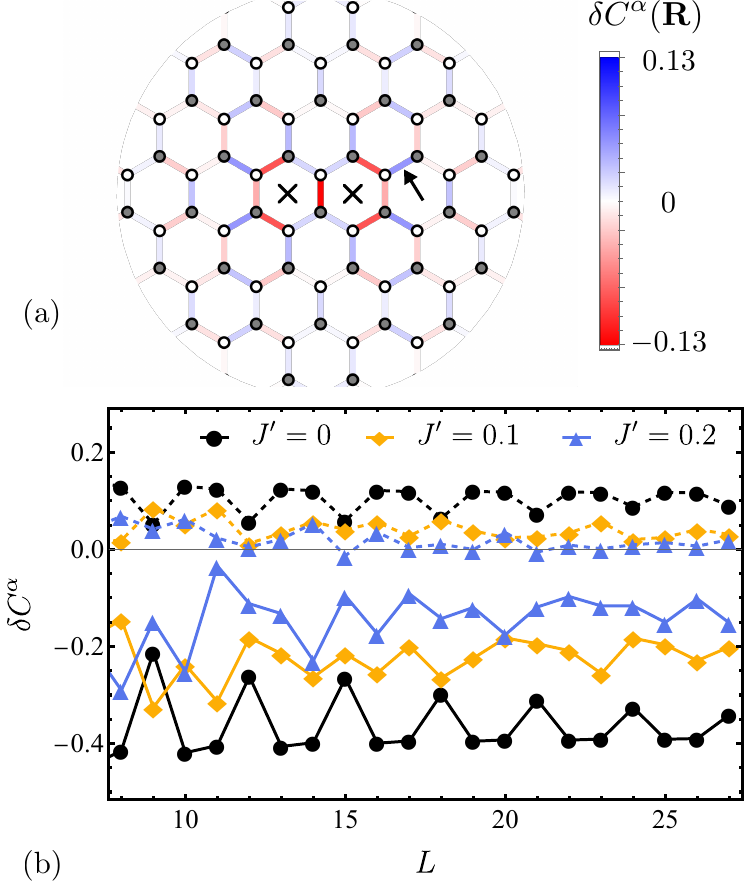}
    \caption{(a) Spatial pattern of nearest-neighbor spin correlations in the vicinity of a vortex pair in the perturbed KHM.  The vortices are created in the plaquettes marked by crosses.   The color scale represents the relative change $\delta C^\alpha(\mb R)$ defined in Eq. (\ref{deltaC}).  Here we use  $J'/J=0.2$ and $L=13$. (b)  \new{ Relative change in the spin  correlations  as a function of system size for three different values of $J'$. The symbols connected by solid lines refer to the $C^z$ correlation on the central $z$ bond between the  visons, which is suppressed in the presence of the visons ($\delta C^z<0$). The dashed line represents the $C^x$ correlation on the bond indicated by an arrow in (a).} } 
    \label{fig:correlation}
\end{figure}

Spin correlations in the ground state of the KHM can be calculated using the conservation of the $\mathbb Z_2$ bond operators \cite{baskaran2007exact}. \new{As a consequence of these conserved quantities, spin-spin correlations vanish identically beyond nearest-neighbor sites. } The nonzero  correlations on a nearest-neighbor bond $\langle i,j\rangle_\alpha$ is given by  \be
C^\alpha_{ij}= \langle \sigma^\alpha_i  \sigma^\alpha_j\rangle =-\ii u_{ij}^\alpha \langle c_ic_j\rangle. 
\ee
In the flux-free sector, the correlation is uniform, and we denote the corresponding value by $C_0$, with $C_0>0$ for ferromagnetic Kitaev interactions. In the two-vison sector, the correlation varies with the  position relative to  the unit cell $\mb R=\mb 0$ where we   flip the  sign of $u^z_{ij}$. Let us denote by $C^\alpha(\mb R)$ the correlation between the spin in sublattice A of unit cell $\mb R$ and its nearest neighbor  on an $\alpha$ bond. 
From the diagonalization of the Hamiltonian on finite lattices, we find that the   correlation $C^z (\mb 0)$ at the central $z$ bond is suppressed upon the creation of the vison pair, whereas other bonds can show enhanced correlations.  We define the relative change in the spin correlation  as 
 \be
\delta C^\alpha(\mb R)=\frac{C^\alpha (\mb R)-C_0}{C_0}. \label{deltaC}
\ee 
The spatial pattern is illustrated  in Fig. \ref{fig:correlation}(a). Anisotropic oscillations in $\delta C^\alpha(\mb R)$ are already present for $J'=0$, but they become more extended in the presence of a Majorana Fermi surface.  This  slow decay of the response with the distance from the vison pair accounts for  the strong finite-size effects in the calculation of observables such as the \new{two-vison gap}. \new{While this spatial pattern is reminiscent of Friedel-like oscillations  expected in gapless spin liquids \cite{lai2011power}, the system sizes accessible in our calculations do not allow us to  determine conclusively whether the correlations contain a component oscillating with wavevector  $2k_F$.}

In Fig. \ref{fig:correlation}(b), we show  the change in the spin correlation as a function of system size for different values of $J'$. \new{The results  for the central $z$ bond are represented by symbols connected by solid lines.}  For $J'=0$, the correlation shows a smooth size dependence if we separate systems sizes  $L \text{ mod }3=0$ from   $L \text{ mod }3\neq0$. In contrast, the results for $J'\neq 0$ show   oscillations analogous to those observed in the \new{two-vison gap}. Interestingly, the magnitude of $\delta C^z(\mb 0)$ decreases as $J'$  increases. \new{The same trend is observed for other correlations, for example on the $x$ bond indicated by an arrow in Fig.  \ref{fig:correlation}(a); see the symbols connected by dashed lines in Fig. \ref{fig:correlation}(b).} This behavior shows that the Majorana Fermi surface tends to ``screen'' the effect of the vison impurity-like   potential on local spin correlations.

\section{Discussion\label{sec:concl}}


We discussed   gapped excitations associated with   the creation of  a pair of adjacent visons in $\mathbb Z_2$ spin liquids with a Majorana Fermi surface. We considered two exactly solvable models, namely a perturbed Kitaev honeycomb model and the CYF kagome model.  In both models,   the Majorana Fermi surface is formed by tuning a parameter that moves a Dirac point in the fermionic dispersion away from zero energy, a mechanism analogous to electron doping or pseudoscalar potentials  in graphene and Dirac semimetals.  

We have shown that the \new{two-vison gap} exhibits oscillations as a function of system size when the low-energy modes define  incommensurate  wave vectors, such as the characteristic size  of the Majorana Fermi surface or the momentum of  a Dirac point located away from high-symmetry points in the Brillouin zone. If one is interested in the \new{two-vison gap} in the thermodynamic limit, the analytical approach that relates the \new{two-vison gap} to scattering phase shifts   becomes instrumental in this case. We have demonstrated an application of this approach by generalizing it   to the CYF model, which contains multiple Majorana bands. On the other hand, the strong finite-size effects should be taken into account when analyzing  properties of  such gapless $\mathbb Z_2$ spin liquids in finite geometries.

We found that the \new{two-vison gap} is maximum when the Dirac point lies at zero energy and   decreases as the Majorana Fermi surface increases.  This effect can be interpreted by noting that, similarly to how metals screen charged  impurities,  the Majorana Fermi surface suppresses the change in   spin correlations near the visons. In a simple picture, the system behaves as if  the dressed  Majorana-vison scattering potential becomes effectively weaker. Since the \new{two-vison gap} is equivalent to the energy cost for introducing the local potential and given by an integral of the exact scattering phase shift, increasing the Majorana Fermi surface tends to lower the \new{two-vison gap}. 

\new{The softening of the local  two-vison gap  uncovered in this work  may be useful in assessing the stability of spin liquid phases with Majorana Fermi surfaces. However, it is important to note that the energy gap for separated visons can differ significantly from that for adjacent visons, particularly in this case where   gapless Majorana modes  mediate algebraically decaying, oscillatory  interactions between   visons \cite{lahtinen2011interacting}. In the solvable models considered in this work, the phase transitions are first order and correspond to level
crossings with other quantum spin liquid ground states.  More generally, in the presence of additional perturbations, the leading instability   driving the transition may   be associated with the condensation of vison pairs or isolated visons \cite{Zhang2021}, whichever has lower energy, or with the  competition between different periodic flux sectors \cite{Zhang2019}.}

 \new{An important open problem is to investigate how the energy gap varies with the distance between the visons. Importantly, finite-size calculations in this case may be strongly affected by long-range vison-vison interactions.} For this purpose, it would be important to  extend the analytical technique to calculate the \new{two-vison gap} for spatially separated visons in the thermodynamic limit.   In addition, while we have focused on a regime of nearly circular Majorana Fermi surfaces, we   note that the CYF model also harbors states with line Fermi surfaces with zero curvature. In this case, the Majorana fermion Green's function shows a rather different spatial dependence  \cite{bauer2019symmetry,Oliviero2024}, which may lead to qualitatively different behavior in the \new{two-vison gap} and local spin correlations. Finally,  it would be interesting to study dynamical response functions, as the Majorana Fermi surface may give rise to power-law threshold singularities in close analogy with the orthogonality catastrophe in metals.

\new{\emph{Data availability.---}The data that support the findings of this study were generated by numerical simulations. The source code and parameters used to generate the simulations for the CYF model is publicly  available at Zenodo \cite{Zenodo}.}

\begin{acknowledgments}
We thank R. Egger for discussions. C.V.S.S. acknowledges funding by Coordena\c{c}\~{a}o de Aperfei\c{c}oamento de
Pessoal de N\'{i}vel Superior   (CAPES). This work was supported by a grant from the Simons Foundation (Grant No. 1023171, R.G.P.), by Finep (Grant No. 1699/24 IIFFINEP, R.G.P.), and Conselho Nacional de Desenvolvimento Cient\'{i}fico e Tecnol\'{o}gico (Grants No. 309569/2022-2 and  404274/2023-4, R.G.P.). 
\end{acknowledgments}

\appendix

\bibliography{refs}

\end{document}